\documentclass[twocolumn,fp]{jpsj3}
\usepackage{graphicx}
\usepackage{txfonts,bm,revsymb}
\usepackage{bm}
\usepackage{color}

\newcommand{\cop}{\hat{c}}
\newcommand{\Hop}{\hat{H}}
\newcommand{\Psiop}{\hat{\Psi}}

\newcommand{\calH}{\mathcal{H}}
\newcommand{\calP}{\mathcal{P}}
\newcommand{\calQ}{\mathcal{Q}}
\newcommand{\calR}{\mathcal{R}}
\newcommand{\calS}{\mathcal{S}}
\newcommand{\calO}{\mathcal{O}}
\newcommand{\calV}{\mathcal{V}}

\newcommand{\veck}{\bm{k}}
\newcommand{\vecell}{\bm{\ell}}

\begin{document}

\title{Topological $s$-wave pairing superconductivity with
spatial inhomogeneity: Mid-gap-state appearance and
robustness of superconductivity
}

\author{Yuki Nagai, Yukihiro Ota, and Masahiko Machida}%
\inst{%
CCSE, Japan  Atomic Energy Agency, 178-4-4, Wakashiba, Kashiwa, Chiba, 277-0871, Japan}%

\abst{
We study the quasiparticle spectrum of 2D topological $s$-wave
superconductors with the Zeeman magnetic field and the Rashba spin-orbit
coupling in the presence of spatial inhomogeneity. 
Solving the real-space Bogoliubov-de Gennes equations, we focus on the
excitations within the superconducting gap amplitude, i.e., the
appearance of mid-gap states. 
Two kinds of potential functions, line-type (a chain of
impurities) and point-type (a single impurity)
ones are examined to take spatial inhomogeneity into account.  
The line setting shows a link of the mid-gap states with the
gapless surface modes indicated by the bulk-boundary correspondence in
topological superfluid. 
The point one shows that the quasiparticles with mid-gap energy
are much easily excited by an impurity when the Zeeman
magnetic field increases within the topological number to be unchanged. 
Thus, we obtain insights into the robustness of a topological
superconductor against non-magnetic impurities. 
Moreover, we derive an effective theory applicable to high magnetic fields. 
The effective gap is the mixture of the chiral $p$-wave and
$s$-wave characters.  
The former is predominant when the magnetic field increases. 
Therefore, we claim that a chiral $p$-wave character of the effective
gap function creates the mid-gap states. 
}

\maketitle

\section{Introduction}
One of the fundamental issues in superconductivity is to study impurity
effects. 
A statement about the response to non-magnetic impurities,
so-called Anderson's theorem plays a central role in this issue; no
significant $T_{\rm c}$ reduction by non-magnetic impurities occurs in
$s$-wave
superconductivity\cite{Anderson,Tsuneto:1962,deGennes:1999,Kopnin:2001,Balatsky}.  
In other words, the impurities produce no additional pole 
in the Green's function within energy scale characterized by
superconducting gap amplitude. 
Thus, when the impurities lead to the decrease of $T_{\rm c}$, the
reduction can be related to poles inside gap energy, mid-gap states
(MGSs).\cite{Hirschfeld;Woelfe:1986,SchmittRink;Varma:1986,Hotta:1993,Preosti;Muzikar:1996,Maki;Puchkaryov:1999,Maki;Puchkaryov:2000}

Studying topological
superconductors\cite{Fu;Berg:2009,Sato:2010,Sato;Fujimoto:2009,Sato;Fujimoto:2010}
is an exciting research topic in condensed matter physics now. 
The bulk-boundary correspondence\cite{Fradkin:2013} in
topological materials implies the presence of gapless (zero-energy) 
modes around defects, as well as on surfaces.
The gapless modes around defects in topological superconductors
are studied in detail.\cite{Read;Green:2000,Teo;Kane:2010,Nishida:2010} 
However, the impurity effects, especially the decrease of $T_{\rm c}$
are not discussed fully, although both impurities and defects induce
spatial inhomogeneity into bulk systems. 

Nagai \textit{et al.}~\cite{Nagai;Machida:1312.3065} studied the
non-magnetic impurity effects in a
2D topological superconductor with on-site $s$-wave pairing, using a
self-consistent $T$-matrix approach.  
No considerable $T_{\rm c}$ decrease is found in a weak
Zeeman magnetic field.
However, the decrease via non-magnetic impurities is
significant in a certain case. 
The reduction is more pronounced when the value of a
topological number increases; a state with a non-zero even
topological number significantly suffers from
non-magnetic impurities, like an unconventional superconducting state. 
Since this topological number is related to the number of the
gapless surface modes,\cite{Sato;Fujimoto:2010} the authors infer from
this result that the robustness of this topological $s$-wave pairing
superconductor against non-magnetic impurities is related to the gapless
surface modes. 
Therefore, it is worth asking how the MGSs induced by an impurity are
related to the gapless surface modes, and how the
unconventional feature appears in $s$-wave pairing. 

In this article, we study the quasiparticle excitations in a 
topological superconductor with spatial inhomogeneity. 
We focus on a 2D topological superconducting
model,\cite{Sato;Fujimoto:2009,Sato;Fujimoto:2010} with on-site $s$-wave
pairing, the Rashba spin-orbit coupling, and the Zeeman magnetic field. 
The model for the inhomogeneity is made by adding a local
potential term to the uniform system. 
We examine two kinds of the setting, line-type and
point-type potentials.  
The former is regarded as a straight chain of non-magnetic impurities,
while the latter is a single one. 

Changing the Zeeman magnetic field and the potential height, the
quasiparticle spectrum is calculated numerically. 
Throughout this article, we focus on a topological phase in which the
topological number is $1$. 
We find that the both potential models induce the MGSs when the
system lies in the topological phase. 
The resultant states are locally bound in the vicinity of
the potential.

We obtain insights into the robustness of the topological superconducting
state against non-magnetic impurity, depending on potential geometry. 
The line setting definitely shows a continuous connection of
the MGSs into the zero-energy bound states, changing the potential into
a hard-wall one. 
Thus, the MGSs in the present model have a link with the gapless surface
modes indicated by the bulk-boundary correspondence, in a certain
setup. 
The point setting indicates that the quasiparticles with mid-gap energy
are much easily excited by impurities when the Zeeman
magnetic field increases within the topological phase. 
We find that this result is consistent with the prediction
on the decrease of $T_{\rm c}$ in Ref.~\citen{Nagai;Machida:1312.3065}.

To understand the numerical results, we discuss why the potential terms 
produce the MGSs. 
From the results in the point setting, we consider that an effective
theory to describe a high-magnetic-field case is useful for the
arguments. 
Thus, we derive an effective gap function in the system without the
potential term, by perturbation about the inverse of the
magnetic-field strength. 
We find that the effective gap is the mixture of the chiral
$p$-wave and $s$-wave components. 
The former becomes primary in the effective gap when the magnetic field
increases. 
Hence, we claim that a chiral $p$-wave feature in the effective gap
function creates the MGSs.

The article is organized as follows. 
In Sec.~\ref{sec:motivation}, we explain our motivation, mainly focusing
on the previous theoretical studies about the non-magnetic impurity
effects in the $s$-wave superconductivity. 
In Sec.~\ref{sec:model}, we explain a model of the 2D topological
$s$-wave superconductors.  
The quasiparticle spectrum for line-type potential is shown in
Sec.~\ref{sec:results1}, and subsequently the results for the
point-type one are shown in Sec.~\ref{sec:results2}. 
In Sec.~\ref{sec:discussion}, we shows the discussions to understand
better the numerical results. 
Specifically, we derive an effective gap function with both the
chiral $p$-wave and $s$-wave characters in Sec.~\ref{subsec:low-energy}. 
In Sec.~\ref{sec:summary}, the summary is shown. 

\section{Motivation and aim}
\label{sec:motivation}
The main issue in this article has a link with the impurity effects in
superconductivity. 
To clarify our motivation, we summarize several theoretical results about
superconducting alloys. 
Then, we show the aim in this article, focusing on our previous
results about non-magnetic impurity
effects.\cite{Nagai;Machida:1312.3065}

\subsection{Anderson's theorem}
A well-known result in the impurity effects is Anderson's
theorem\cite{Anderson}. 
The critical temperature in $s$-wave superconductors does not
significantly decrease by non-magnetic impurities. 
The idea is to focus on Kramers-degenerate electrons. 
If a metal has the time-reversal symmetry, the instability leading to 
superconductivity comes from the time-reversed pairs of electrons
on the Fermi surface. 
Then, the perturbation not splitting the Kramers
degeneracy has no effect on $T_{\rm c}$.

It is an important issue in superconductivity to ask how $T_{\rm c}$
decreases in the presence of impurities when the conditions in the
above argument are changed.  
Several ways of changing the conditions are available. 
One way is to change the superconducting character, keeping the
time-reversal property of the normal electrons. 
Another way is to violate the time-reversal symmetry, with
$s$-wave pairing. 
Considering spin-orbit couplings can also lead to the change of the
conditions, since their absence is implicitly assumed
in the typical arguments of Anderson's theorem.\cite{deGennes:1999}

\subsection{Mid-gap-state appearance in various settings}
Since Anderson's theorem includes an argument of $T_{\rm c}$, the
quantitative discussion is needed for studying the validity in various
physical settings. 
Studying the response to magnetic impurities gives us a clue for this
issue. 
Magnetic impurities lead to the decrease of $T_{\rm c}$ in $s$-wave
superconductors.\cite{deGennes:1999,Balatsky} 
This is associated with the formation of the
bound states around the impurities, with mid-gap energy 
$E \in (-\Delta,\Delta)$, where $\Delta$ is the superconducting gap
amplitude. 
The mid-gap-state appearance corresponds to the occurrence of additional
poles in the Green's functions. 
The modification of the anomalous Green's functions can change 
$T_{\rm c}$.

We summarize the theoretical results on the impurity effects in the
$s$-wave superconductivity, relevant to our issue.

\subsubsection{Magnetic impurities}
The presence of magnetic impurities leads to significant decrease of 
$T_{\rm c}$.\cite{deGennes:1999,Balatsky}
This is related to the appearance of additional poles in the
Green's functions, as pointed out above. 

\subsubsection{Zeeman magnetic field and non-magnetic impurities}
Zeeman magnetic field breaks time reversal symmetry. 
One can expect that an $s$-wave superconducting state in the presence of
the Zeeman magnetic field 
is not robust against non-magnetic impurities. 
The expectation is partially correct, but not always holds. 
We will argue this point below. 

Ohashi\cite{Ohashi} proposed a way of making a magnetic
impurity in superfluid fermionic gases and showed the formation of bound
states in an $s$-wave state. 
The keys of creating a magnetic impurity are to make Zeeman field by
the population imbalance between atomic components and to use so
strong on-site pairing potential that the Hartree term is predominant. 
Thus, the interplay between a non-magnetic impurity and the Zeeman field
leads to the mid-gap-state appearance, under strong on-site pairing
potential. 
However, in a weak-coupling superconducting theory the
Hartree term is negligible. 

Using a weak-coupling approach, Sau and Demler\cite{Sau;Demler:2013}
studied the formation of bound states by impurities in a semiconductor
nanowire in the 
proximity to an $s$-wave superconductor, similar to our 
model; both the Zeeman field and the spin-orbit coupling are involved.   
They point out that no MGS bound to non-magnetic impurities
is found when the magnetic field takes a non-zero value, but the
spin-orbit coupling vanishes. 
The form of the Green's functions in the system without
impurities supports this statement. 
One can find that when only the Zeeman magnetic field exists, the
Green's functions are expressed by a block diagonal matrix with two
sectors, each of which is composed of the ones in the
Bardeen-Cooper-Schrieffer theory. 
The Zeeman magnetic field gives an energy shift in the Green's
functions.\cite{Ichioka;Machida:2007}
Thus, using the standard arguments that non-magnetic impurity
do not induce the bound states in $s$-wave superconductors, the
assertion is obtained. 

The above arguments indicate that the non-magnetic impurity effects
under the Zeeman magnetic field are not so simply understood by the
magnetic-impurity effects. 
We remark that our target is a weak-coupling $s$-wave superconductor. 

\subsubsection{Spin-orbit couplings and non-magnetic impurities}
Studying specific models with spin-orbit couplings shows a similar
statement to Anderson's theorem. 
Non-magnetic disorder has no effect on the critical temperature in a
non-centrosymmetric superconductor with $s$-wave pairing.\cite{Samokhin} 
In the proximity-induced superconductivity in a semiconductor nanowire
only with the Rashba spin-orbit coupling\cite{Sau;Demler:2013}, no 
additional pole in the Green's function appears by a non-magnetic
impurity.  

\subsection{Impurity effects and a topological number}
The presence of \textit{both} the Rashba spin-orbit 
coupling and the Zeeman magnetic field lead to non-trivial response
to non-magnetic
impurities.\cite{Sau;Demler:2013,Hu;Liu:2013,Nagai;Machida:1312.3065} 
To see this point, we summarize our previous
results\cite{Nagai;Machida:1312.3065} in a 2D $s$-wave topological
superconductor. 
The detail of the model is explained in Sec.~\ref{sec:model}. 

Nagai \textit{et al.}\cite{Nagai;Machida:1312.3065} numerically
calculated $T_{\rm c}$ in the presence 
of non-magnetic impurities, using a self-consistent $T$-matrix approach. 
When the system in the clean limit lies in a non-topological region (i.e.,
a low-magnetic-field region), no significant $T_{\rm c}$
reduction occurs by non-magnetic impurities. 
However, the decrease of $T_{\rm c}$ appears when taking a higher
magnetic field. 
Moreover, the calculations of a topological invariant,
Thouless-Kohmoto-Nightingale-Nijs (TKNN) number\cite{Thouless;Nijs:1982}
indicate the presence of a correlation of
the TKNN number with the $T_{\rm c}$ reduction; 
the reduction is more pronounced when the TKNN number increases. 
We remark that the proximity-induced superconductivity in a
semiconductor nanowire\cite{Sau;Demler:2013} also shows the formation of
a bound state around a non-magnetic impurity, increasing the Zeeman
magnetic field with non-zero spin-orbit coupling. 

\subsection{Settings in the present issue: Potential models}
The above results motivate us to examine how in our model the
mid-gap-state appearance connects with a
topological property of the system. 
We focus on the fact that in the clean-limit bulk system the TKNN number is
related to the number of the gapless surface modes in the system with
edges.\cite{Sato;Fujimoto:2010}  
Thus, it is a meaningful issue to consider the relation of the
mid-gap-state appearance with the gapless surface modes indicated by the
bulk-boundary correspondence. 
Studying the production of bound states [with energy 
$E\in (-\Delta,\Delta)$] in potential models is suitable for this issue. 
Changing potential strength, one can examine the
bound-state production by both a impurity (i.e., finite barrier) and a
defect (i.e., infinitely-high barrier). 

Now, we show the settings in this article. 
We focus on two kinds of potential models, line-type potential and
point-type potential, as seen in Fig.~\ref{fig:fig1}. 
The line-type potential may deform into a rigid wall through 
increase of the strength. 
The setup is also regarded as a straight chain of impurities. 
Thus, we first study the line setting, to understand
the mid-gap-state appearance in terms of the bulk-boundary
correspondence. 
A line segment can be a point when one shortens the
line length continuously. 
Therefore, our next task is to study the point-type potential, as a
limit of the line setting. 
The point setting is related to the calculations of $T_{\rm c}$
with a $T$-matrix approach when the impurity concentration is quite low.  

\section{Model}
\label{sec:model}
The system is described by a tight-binding model on an 
$N_{x}\times N_{y}$ square lattice, with on-site $s$-wave
pairing and the Rashba spin-orbit coupling. 
Moreover, the Zeeman magnetic field is applied along the direction
perpendicular to the $xy$-plane. 
The mean-field Hamiltonian is 
\begin{eqnarray}
 \Hop 
&=& 
-t \sum_{<j,j^{\prime}>}\sum_{\sigma} 
\cop^{\dagger}_{j,\sigma} \cop_{j^{\prime},\sigma}
-
\mu \sum_{j}\sum_{\sigma} 
\cop^{\dagger}_{j,\sigma} \cop_{j,\sigma}
\nonumber \\
&&
-\frac{\alpha}{2}
\sum_{j}
[
(
\cop^{\dagger}_{j-e_{x},\downarrow} \cop_{j,\uparrow}
-
\cop^{\dagger}_{j+e_{x},\downarrow} \cop_{j,\uparrow}
)
\nonumber \\
&&
\quad\quad
+
i
(
\cop^{\dagger}_{j-e_{y},\downarrow} \cop_{j,\uparrow}
-
\cop^{\dagger}_{j+e_{y},\downarrow} \cop_{j,\uparrow}
)
+
\text{h.c.}
]
\nonumber \\
&&
-h\sum_{j} (
\cop^{\dagger}_{j,\uparrow}\cop_{j,\uparrow}
-
\cop^{\dagger}_{j,\uparrow}\cop_{j,\uparrow}
) 
\nonumber \\
&&
+
\sum_{j}
(
\Delta \cop^{\dagger}_{j,\uparrow}\cop^{\dagger}_{j,\downarrow}
+
\text{h.c.}
)
+
\sum_{j}\sum_{\sigma} 
V_{j} \cop^{\dagger}_{j,\sigma}\cop_{j,\sigma},
\label{eq:tb_hamiltonian}
\end{eqnarray} 
with the nearest neighbor hopping matrix element $t\, (>0)$, the chemical
potential $\mu$, the spin-orbit coupling constant $\alpha\,(>0)$, the
magnitude of the Zeeman magnetic field $h$, and the superconducting gap
$\Delta$. 
The lattice constant is normalized by $1$. 
The annihilation and creation operators of electrons with spin
$\sigma\,(=\uparrow,\,\downarrow)$ are, respectively, 
$\cop_{j,\sigma}$ and $\cop^{\dagger}_{j,\sigma}$ on site
$j=(j_{x},\,j_{y})$. 
In the third term, i.e., Rashba spin-orbit coupling, the symbol
$e_{x}$ means the unit vector along $x$-axis: $e_{x}=(1,0)$. 
Similarly, $e_{y}$ is the unit vector along $y$-axis. 
The last term is the potential term. 
We focus on two kinds of the potential functions; (i) line-type
potential ($V_{j} = V^{(\rm L)}_{j}$) and (ii) point-type potential
($V_{j} = V^{({\rm P})}_{j}$), where
\begin{eqnarray}
&&
V^{({\rm L})}_{j} = V \delta_{j_{x},N_{x}/2}, 
\label{eq:line_potential} \\ 
&&
V^{({\rm P})}_{j} = V \delta_{j_{x},N_{x}/2}\, \delta_{j_{y},N_{y}/2}.
\label{eq:point_potential}
\end{eqnarray}
The potential height $V$ is positive (repulsive potential). 
We will use the Fourier-transformed model along $y$-axis, for case (i).  
Tight-binding Hamiltonian (\ref{eq:tb_hamiltonian}) is equivalent to a 2D topological
superconducting model in ultra-cold atomic
gases~\cite{Sato;Fujimoto:2009,Sato;Fujimoto:2010}, if the potential
term is absent. 
Our model also describes a proximity-induced superconducting system
on the interface between semiconductors and superconductors. 
In the junction systems, the potential term can be
implemented by controlling the chemical potential via local
gate voltage.\cite{Alicea;Fisher:2011} 

Let us concisely summarize the topological properties of our system in
clean limit ($V_{j}=0$ for any $j$). 
The spatial uniformity allows us to write down the tight-binding
Hamiltonian in momentum space. 
The electron annihilation and creation operators are, respectively, 
\(
\cop_{\veck,\sigma}
\)
and
\(
\cop^{\dagger}_{\veck,\sigma}
\). 
They are bound as the $4$-component vectors, 
\(
\Psiop_{\veck} 
=
\!^{\rm t} (
\cop_{\veck,\uparrow},\,
\cop^{\dagger}_{-\veck,\uparrow},\,
\cop_{\veck,\downarrow},\,
\cop^{\dagger}_{-\veck,\downarrow} )
\)
and 
\(
\Psiop_{\veck}^{\dagger} 
=
(
\cop^{\dagger}_{\veck,\uparrow},\,
\cop_{-\veck,\uparrow},\,
\cop^{\dagger}_{\veck,\downarrow},\,
\cop_{-\veck,\downarrow} )
\). 
We find that 
\(
\Hop |_{V=0}
= 
(1/2)
\sum_{\veck}\Psiop_{\veck}^{\dagger} \calH_{\veck} \Psiop_{\veck}
\), with the Bogoliubov-de Gennes (BdG) Hamiltonian, 
\begin{eqnarray}
 \calH_{\veck}
&=&
 \frac{s^{0} + s^{3}}{2} \otimes (\varepsilon_{\veck} - h) \tau^{3}
+
 \frac{s^{0} - s^{3}}{2} \otimes (\varepsilon_{\veck} + h) \tau^{3}
\nonumber 
\\
&&
+ s^{1} \otimes \alpha \ell_{1,\veck} \tau^{0} 
+ s^{2} \otimes \alpha \ell_{2,\veck} \tau^{3} 
\nonumber \\
&&
+ s^{2} \otimes ( i\Delta \tau^{+} + \text{h.c.}),
\label{eq:bdg_hamiltonian}
\end{eqnarray} 
where 
\(
\varepsilon_{\veck} = -\mu - 2t (\cos k_{x} + \cos k_{y})
\), 
\(
\ell_{1,\veck} = \sin k_{y}
\), 
and 
\(
\ell_{2,\veck} = - \sin k_{x}
\). 
The $i$th component of the $2 \times 2$ Pauli matrices ($i=1,\,2,\,3$)
is written by $s^{i}$ for spin. 
Similarly, the Pauli matrices for the Nambu space are written by
$\tau^{i}$.  
The $2 \times 2$ identity matrices for spin and the Nambu space are,
respectively, $s^{0}$ and $\tau^{0}$. 
The ladder operators in the Nambu space are defined by 
\(
\tau^{\pm} = (1/2)(\tau^{1} \pm i \tau^{2})
\).  
The diagonalization of Eq.~(\ref{eq:bdg_hamiltonian}) leads to the bulk
spectrum 
\begin{align}
E(\bm{k}) &= \sqrt{
\varepsilon_{\bm{k}}^{2}+\alpha^{2} |\bm{\ell}_{\bm k}|^{2}+h^{2} 
+ |\Delta|^{2} \pm 2 \xi_{\bm{k}} 
}, \label{eq:bulk}
\end{align}
with 
\(
\bm{\ell}_{\bm{k}}
=
(\ell_{1,\bm{k}},\, \ell_{2,\bm{k}}) 
\) and 
\(
\xi_{\bm{k}}  = \sqrt{
\varepsilon_{\bm{k}}^{2} \alpha^{2} |\bm{\ell}_{\bm k}|^{2} 
+ (\varepsilon_{\bm{k}}^{2} + |\Delta|^{2}) h^{2}}
\). 
The topological phase
transition occurs when the energy gap of bulk spectrum (\ref{eq:bulk}) 
closes at specific points in the Brillouin zone.~\cite{Sato;Fujimoto:2010} 
According to Table I in Ref.~\citen{Sato;Fujimoto:2010}, 
the topological property of the superconducting state for $\mu > 2t $
and $h < 3t$ changes from a trivial
phase to a non-trivial one, when $h$ is greater than a critical value, 
\begin{align}
h > \sqrt{(4t - \mu)^{2} + \Delta^{2} }. \label{eq:tsc}
\end{align}

\section{Results}
We numerically diagonalize tight-binding Hamiltonian
(\ref{eq:tb_hamiltonian}), to examine the quasiparticle excitations in
this system. 
The pair potential is $\Delta = 0.35t$ and the
chemical potential is $\mu = 3.5t$, throughout this article. 
We accurately evaluate the eigenvalues and the eigenstates within
$(-\Delta,\Delta)$, with a method by Sakurai and
Sugiura~\cite{Sakurai;Sugiura:2003}. 
The authors in Ref.~\citen{Sakurai;Sugiura:2003} propose a practical
way of solving a generalized eigenvalue problem, with a
restricted eigenvalue domain. 
This approach leads to a numerical construction of a small-size
effective Hamiltonian, keeping the essential properties of the
original large-size Hamiltonian. 
The central idea is an efficient calculation of the
projector onto an aimed energy domain. 
The contour integral for the projector on complex plain is approximated 
by numerical quadrature. 
The application to different superconducting issues and the detail
implementation way are shown in Ref.~\citen{Nagai;Sakurai:2013}. 

\subsection{Line-type potential}
\label{sec:results1}
\begin{figure}[thb]
\begin{center}
\begin{tabular}{p{ \columnwidth}} 
\resizebox{ \columnwidth}{!}{\includegraphics{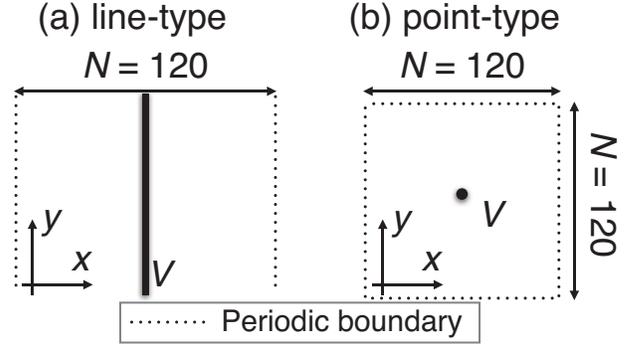}} 
\end{tabular}
\end{center}
\caption{\label{fig:moshiki} 
Schematic diagrams of a system with (a) line-type potential and (b) point-type potential.
} 
\end{figure}

We concentrate on line-type potential (\ref{eq:line_potential}). 
The set up is shown in Fig.~\ref{fig:moshiki}(a). 
The translational symmetry along $y$-axis leads to a $k_{y}$-resolved
Hamiltonian from Eq.~(\ref{eq:tb_hamiltonian}). 
The number of the lattice size along $x$-axis is $N_{x} = 120$. 
We impose the periodic boundary condition along $x$-axis. 

\begin{figure}[thb]
\begin{center}
\begin{tabular}{p{ \columnwidth}} 
\resizebox{ \columnwidth}{!}{\includegraphics{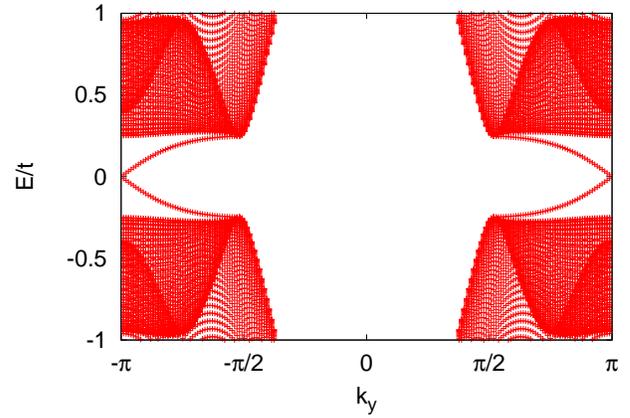}} 
\end{tabular}
\end{center}
\caption{\label{fig:fig1} 
(Color online) 
Quasiparticle spectrum for a ultra-high line-type potential function $(V = 100t)$, along $y$-axis. 
See Fig.~\ref{fig:moshiki}(a), as well. The pair potential is $\Delta = 0.35t$, and the chemical potential 
is $\mu = 3.5t$. 
Since we set the Zeeman magnetic field as $h = 1t$, the system without the potential stays at a topological phase 
[See, Eq.~(\ref{eq:tsc}]. The gapless states appear at $k_{y}=\pm \pi$. }
\end{figure}

We study about the case of the Zeeman-magnetic-field strength to be
$h=t$, for a while.  
We find that, from Eq.~(\ref{eq:tsc}), a topological 
superconducting state occurs in the clean-limit bulk system. 
Figure \ref{fig:fig1} shows that the potential with $V=100\, t$ induces the 
almost-zero-energy states. 
Since the system without the potential is in a topological phase, 
these zero-energy states are regarded as the gapless bound states on
the surfaces of the potential wall. 
In other words, the line-type potential with $V=100\,t$ can be regarded as a line defect,
i.e., wall. 
This ultra-high potential barrier completely divides the system into the two
subsystems, left and right parts; the left part is separated from the
right part, by vacuum. 

%

\begin{figure}[thb]
\begin{center}
\begin{tabular}{p{0.33 \columnwidth} p{0.33 \columnwidth} p{0.33 \columnwidth}} 
\resizebox{0.35 \columnwidth}{!}{\includegraphics{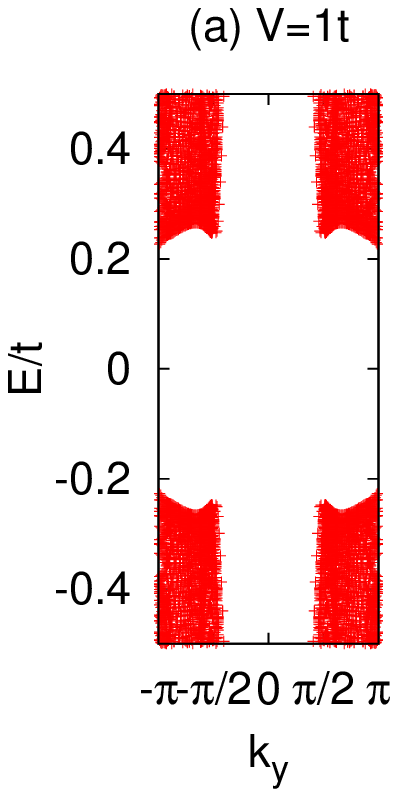}} &\resizebox{0.35 \columnwidth}{!}{\includegraphics{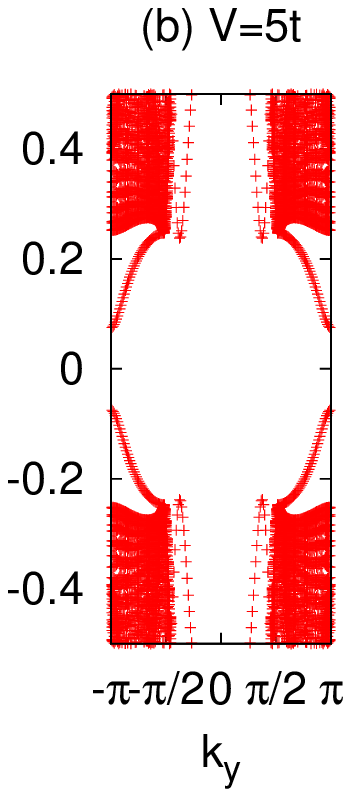}}  &\resizebox{0.35 \columnwidth}{!}{\includegraphics{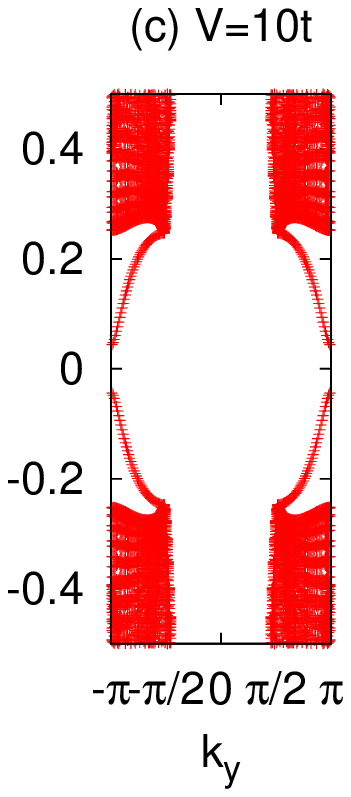}} 
\end{tabular}
\end{center}
\caption{\label{fig:fig2} 
(Color online) 
Quasiparticle spectrum for line-type potential, with different barrier heights. 
The other physical parameters are the same as in Fig.~\ref{fig:fig1}. 
When $V \ge 5 t$, the mid-gap states appear.}
\end{figure}

\begin{figure}[thb]
\begin{center}
\begin{tabular}{p{ \columnwidth}} 
\resizebox{ \columnwidth}{!}{\includegraphics{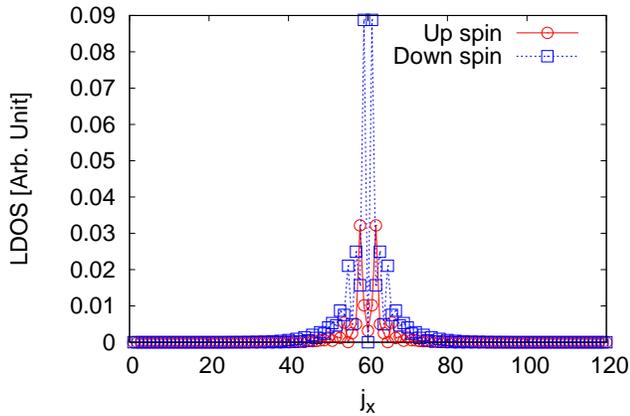}} 
\end{tabular}
\end{center}
\caption{\label{fig:fig4} 
(Color online) 
Spin-resolved local density of states, at energy $E = 7.4559 \times 10^{-2}t$, for line-type potential with 
height $V = 5t$. See Fig.~\ref{fig:fig2}(b), as well. 
The other parameters are the same as in Fig.~\ref{fig:fig1}. }
\end{figure}

The result for the ultra-high-barrier line-type potential motivates us
to study lower-barrier cases.  
Figure \ref{fig:fig2} shows the energy dispersions, with 
$V=1\,t$, $5\,t$, and $10\,t$. 
We find that for $V = 5\,t$ and $10\,t$ the quasiparticle excitations
occur, within the energy domain $(-\Delta,\,\Delta)$. 
Thus, the line-type potentials with either high or intermediate barriers 
induce the MGSs. 
Changing $V$ means the continuous deformation of the energy spectrum of
the system. 
Therefore, these MGSs can continuously connect with the
zero-energy states, when $V$ increases. 
Another evidence for the MGSs to be related to the
zero-energy states is obtained, calculating the local density of states
(LDOS). 
Figure \ref{fig:fig4} shows the local density of states at 
energy $E=7.4559\times 10^{-2}\,t$, for $V=5\,t$. 
This energy corresponds to the lowest absolute eigenvalue of the
Hamiltonian. 
We find that the corresponding states are located on the surfaces of the
potential wall. 
The spin-imbalanced behavior in the LDOS attributes to the Zeeman
magnetic field and the chemical-potential value; the first and the
second terms in Eq.~(\ref{eq:bdg_hamiltonian}) imply that the
down-spin component mainly contributes to the LDOS near zero energy.  

%
%

\begin{figure}[t]
\begin{center}
\begin{tabular}{p{ \columnwidth}} 
(a)\resizebox{ \columnwidth}{!}{\includegraphics{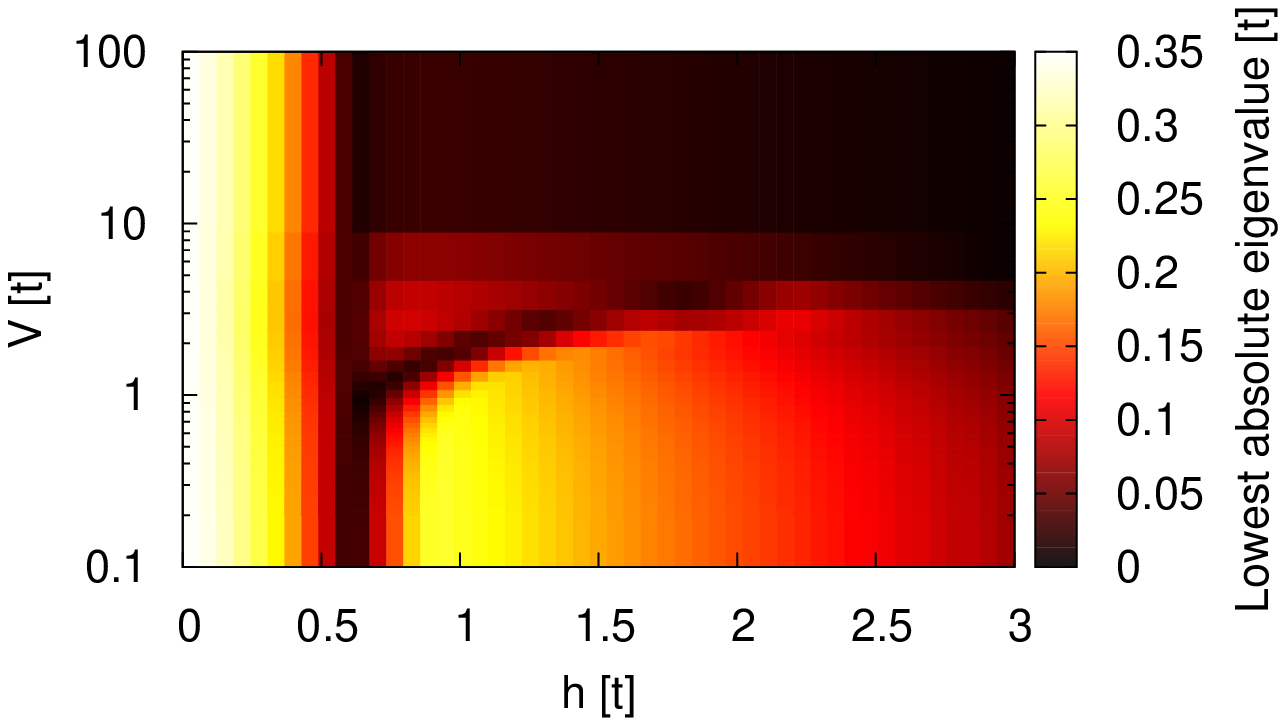}} \\
(b)\resizebox{ 0.9\columnwidth}{!}{\includegraphics{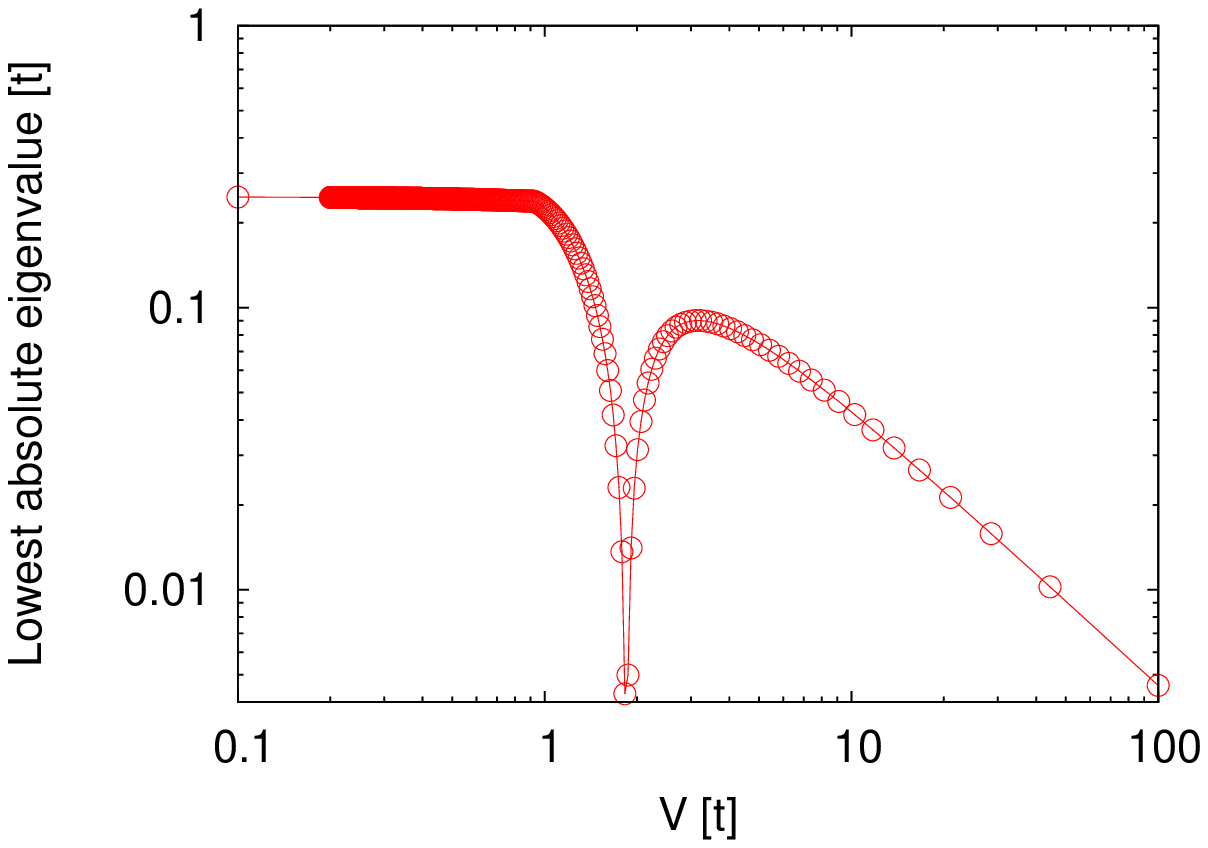}} 
\end{tabular}
\end{center}
\caption{\label{fig:fig5} 
(Color online) 
(a) Eigenvalue of tight-binding Hamiltonian (\ref{eq:tb_hamiltonian}),
 with the lowest absolute value, for line-type potential. 
The magnetic field $h$ (the horizontal axis) and the potential height
 $V$ (the vertical axis, with logarithmic scale) are changed. 
We set the pair potential $\Delta = 0.35t$ and the chemical potential
 $\mu = 3.5t$. 
The black color means the existence of the zero-energy states. 
(b) Section in (a), with a fixed magnetic field strength ($h = 1t$).
The value at $V=0.1t$ corresponds to the bulk spectral gap. Therefore,
 the behavior for $V > 2 t$ indicates the presence of the mid-gap
 states. 
} 
\end{figure}

Now, we turn into a systematic study of the quasiparticle spectrum under
the line-type potential, varying $h$ and $V$. 
Figure \ref{fig:fig5}(a) shows a $h$-$V$ diagram of the lowest absolute
eigenvalue. 
Figure \ref{fig:fig5}(b) shows a section of the $h$-$V$ diagram, with
$h=1\,t$. 
The critical magnetic field for the topological transition
is $h \approx 0.61\,t$, from Eq.~(\ref{eq:tsc}). 
We focus on the case when $h > 0.61\,t$ (i.e., topological phase). 
For a high-barrier region ($V > 5\,t$), we find the appearance of the
MGSs, which is indicated by the dark-colored area in
Fig.~\ref{fig:fig5}(a). 
Figure \ref{fig:fig5}(b) shows that the energy of the MGSs 
is proportional to $1/V$. 
For a ultra-low barrier region ($V < 1\,t$), the lowest
absolute eigenvalue does not significantly reduce. 
Indeed, we can find that this quantity is equal to the
lowest energy of the bulk states. 
Therefore, no MGS is excited by the ultra-low potential.
We obtain a curious result when $1 \,t < V < 5\,t$. 
The lowest absolute eigenvalue is zero (i.e.~ gap closing) at the
specific potential height, depending on $h$. 
Figure \ref{fig:fig5}(b) also shows this behavior manifestly. 
We show an interpretation for this anomalous 
gap-closing behavior in Sec.~4.1.

\subsection{Point-type potential}
\label{sec:results2}
The results in the previous subsection show that the potential is
regarded as a rigid wall when $V=100\,t$. 
Our next concern is to examine whether the same behavior occurs in
different settings. 
We examine the point-type potential, as a line setting with
ultimately-short line length. 
We mainly show the characters of the lowest absolute eigenvalue of
the Hamiltonian to examine the presence of zero-energy bound states. 

\begin{figure}[t]
\begin{center}
\begin{tabular}{p{0.7 \columnwidth} } 
\resizebox{ 0.7 \columnwidth}{!}{\includegraphics{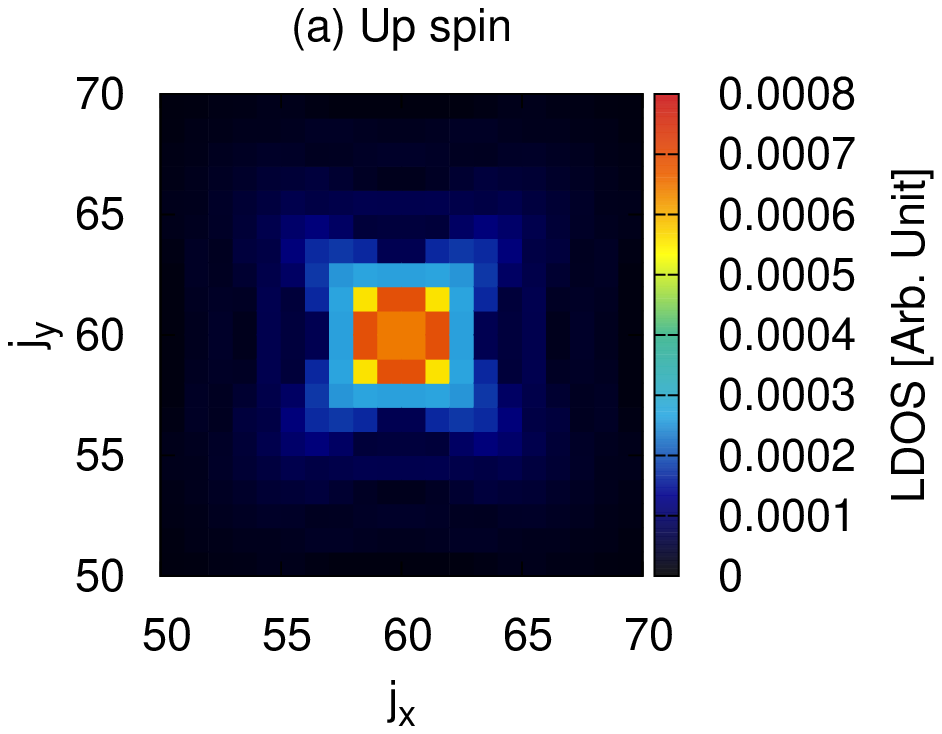}} \\
\resizebox{  0.7 \columnwidth}{!}{\includegraphics{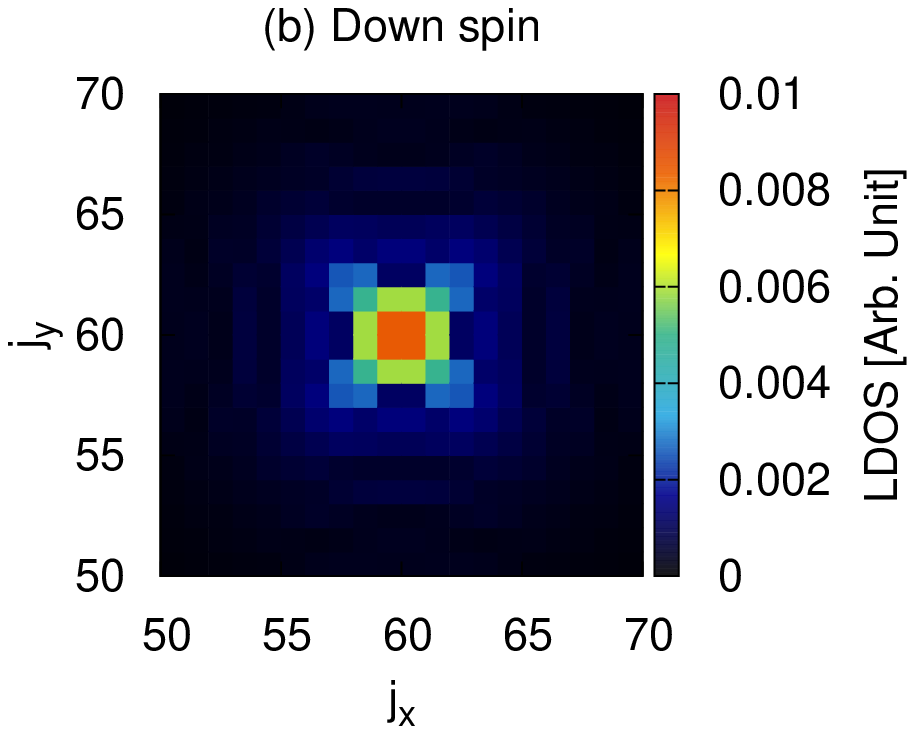}} 
\end{tabular}
\end{center}
\caption{\label{fig:fig7} 
(Color online) Spin-resolved local density of states 
at energy $E =  9.8290 \times 10^{-2}t$ in the system with a point-type potential ($V = 100t$). See, Fig.~\ref{fig:moshiki}(b). 
The Zeeman magnetic field is $h = 2t$. 
The other parameters are the same as in Fig.~\ref{fig:fig1}. 
(a) Up spin component and (b) down spin component of the wave functions. 
We note that in this figure we focus on the area around the potential center.}
\end{figure}

\begin{figure}[thb]
\begin{center}
\begin{tabular}{p{ \columnwidth}} 
\resizebox{ \columnwidth}{!}{\includegraphics{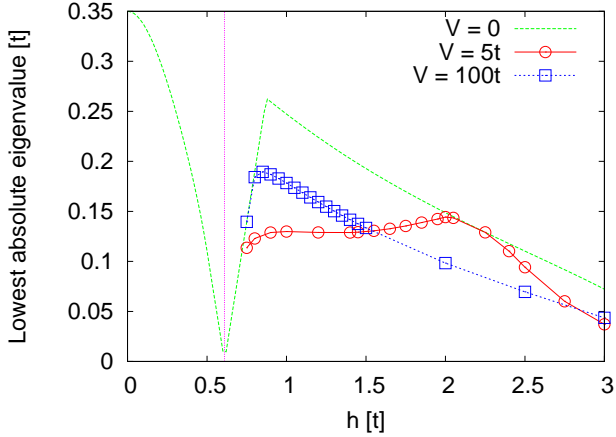}} 
\end{tabular}
\end{center}
\caption{\label{fig:fig8} 
(Color online) 
Magnetic-field dependence of the lowest absolute eigenvalue for point-type potential, with heights $V = 5t$ 
and $V = 100t$. 
The dashed line $(V = 0t)$ is obtained by exact bulk energy spectrum (\ref{eq:tsc}). 
The other parameters are the same as in Fig.~\ref{fig:fig1}. }
\end{figure}

Let us study point-type potential (\ref{eq:point_potential}), as seen in
Fig.~\ref{fig:fig1}(b). 
We set $N_{x}=N_{y}=120$, with the periodic boundary condition. 
Our calculations focus on the case for $h > 0.61\,t$ (topological
phase). 
Figure \ref{fig:fig7} shows the spin-resolved LDOS at energy
$E=9.8290 \times 10^{-2}\,t$, with $h=2\,t$ and $V=100\,t$. 
This energy is the lowest absolute eigenvalue of the Hamiltonian. 
We find that the bound states occur around the point potential, with
non-zero energy. 
Figure \ref{fig:fig8} shows the $h$-dependence of the lowest absolute
eigenvalue of the Hamiltonian, with different heights $V=5\,t$ and 
$100\,t$. 
The curve for $V=0$ is calculated by exact bulk energy
spectrum (\ref{eq:bulk}). 
This curve predominates the plots for $V\neq 0$. 
The behavior for large $V$ (e.g.,$V=100\,t$) is
consistent with the calculations of the MGSs in a similar model to the
present one by Hu \textit{et al.}\cite{Hu;Liu:2013}
The point-type potential leads to the MGSs at the potential center. 
However, the zero-energy bound states are not induced by the potential. 
(See, Table \ref{table:1}). 
We are going to argue this point in Sec.\ref{subsec:non-zero-point}. 

Now, we study the relation of the mid-gap-state appearance to the
decrease of $T_{\rm c}$ in the 2D topological $s$-wave superconductor. 
Since the point-type potential is regarded as the scattering
center of a single non-magnetic impurity, the point setting is 
relevant to the previous $T$-matrix calculations. 

Let us focus on the results in the ultra-high potential barrier (blue
squares) shown in Fig.~\ref{fig:fig8}. 
We denote the lowest absolute eigenvalue as $E^{\rm LA}_{V}$. 
$E^{\rm LA}_{V=0}$ corresponds to the spectrum gap in the
DOS without impurities (the green dashed line in Fig.~\ref{fig:fig8}). 
Moreover, we denote the ratio of $E^{\rm LA}_{V}$ to  
$E^{\rm LA}_{V=0}$ as 
\(
\epsilon^{\rm LA}_{V} \,(=
E^{\rm LA}_{V} \, / E^{\rm LA}_{V=0} 
)
\); 
$\epsilon^{\rm LA}_{V}$ is the lowest
quasiparticle energy normalized by the spectrum gap in the clean limit. 
The data in Fig.~\ref{fig:fig8} indicates that 
$\epsilon^{\rm LA}_{V=100\,t}$ decreases when $h$ increases. 
We can find that 
\(
\epsilon^{\rm LA}_{V=100 t} \approx 0.72
\) for $h=1\, t$, 
\(
\epsilon^{\rm LA}_{V=100 t} \approx 0.66
\) for $h=2\,t$, 
and 
\(
\epsilon^{\rm LA}_{V=100 t} \approx 0.6
\) for $h=3\, t$. 
Thus, for a relatively high magnetic field, the quasiparticles with
mid-gap energy are much easily excited by impurities. 
These behaviors are consistent with our previous
calculations.\cite{Nagai;Machida:1312.3065} 
The decrease of $T_{\rm c}$ is not significant
for $h=1 t$ [Fig.5(a) in Ref.\citen{Nagai;Machida:1312.3065}], while is
pronounced for $h=2 t$ [Fig.5(b) in
Ref.\citen{Nagai;Machida:1312.3065}]. 
Therefore, the location of the MGSs in energy domain can lead to
understanding how the robustness of the present model against impurities
is ruled by the magnetic field. 
Systematic studies, such as dependence on scattering models and full
self-consistent calculations are needed for the precise location of the
poles leading to a significant decrease of $T_{\rm c}$ by a non-magnetic
impurity. 


\begin{table*}[t]
\caption{Difference between line-type and point-type potentials. Anomalous gap-closing is discussed in Sec.~4.1. 
}
\label{table:1}
\begin{center}
\begin{tabular}{lcccl}
\hline
&Line-type potential  &Point-type potential \\
\hline
Bound states ($V \rightarrow \infty$) & 
Zero energy  & 
Non-zero energy \\
Anomalous gap-closing& Yes &No 
\\
\hline 
\end{tabular} 
\end{center}
\end{table*}

\section{Discussions}
\label{sec:discussion}
\subsection{$h$-$V$ diagram for line-type potential}
\begin{figure}[t]
\begin{center}
\begin{tabular}{p{ \columnwidth}} 
(a)\resizebox{ \columnwidth}{!}{\includegraphics{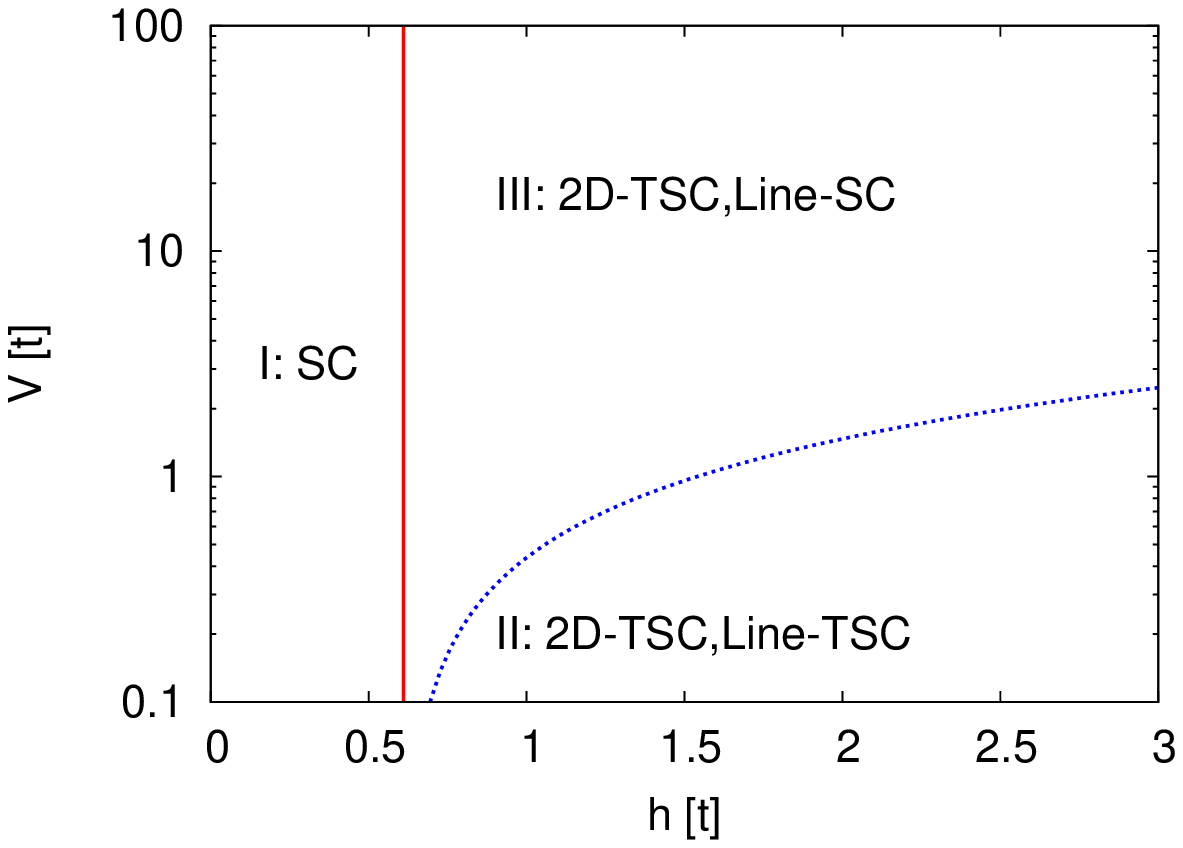}} \\
(b)\resizebox{ \columnwidth}{!}{\includegraphics{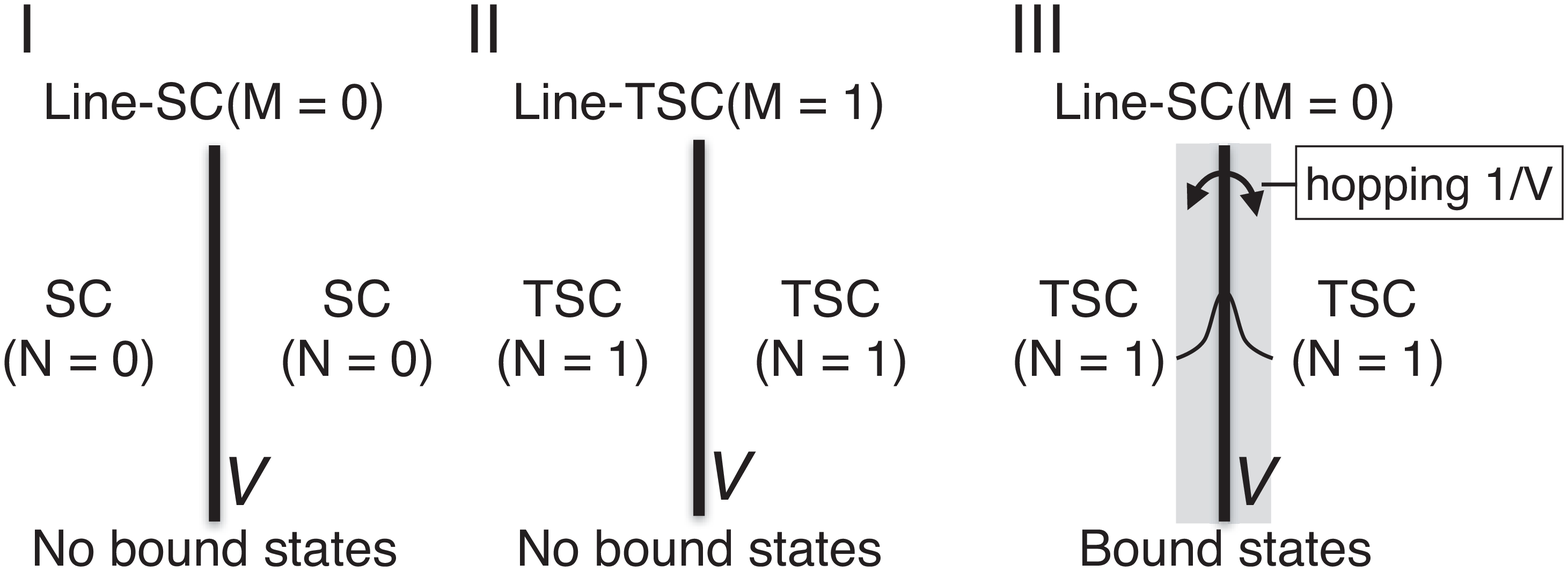}} 
\end{tabular}
\end{center}
\caption{\label{fig:fig6} 
(Color online) (a) Phase diagram with respect to a topological
 invariant. The red solid line denotes the topological phase transition
 in the bulk, defined by Eq.~(\ref{eq:tsc}). 
The blue dotted line denotes the topological phase transition on the
 line-potential, defined by Eq.~(\ref{eq:1dtsc}). 
(b) Schematic diagram of the different phases. The phase III only has
 the bound states at the boundary. } 
\end{figure}

Let us explain the behaviors on the $h$-$V$ diagram shown in
Fig.~\ref{fig:fig5}, in terms of a topological number, 
The topological number for classifying the present 2D topological
superconductor is the TKNN number.~\cite{Thouless;Nijs:1982}
The bulk topological superconducting state for $(3\,t >)\,h > 0.6\,t$ and
$\mu > 2\,t$ has the topological number $N=1$, as seen in
Ref.~\citen{Sato;Fujimoto:2010}. 
Since the potential height is moderate, we may consider a local
property of the topological state \textit{on} the line (effective 1D
superconducting system). 
The effective chemical potential on the line can be $\mu - V$, owing to
the local filling change via the potential. 
According to the typical arguments in a 1D Majorana nanowire in a
semiconductor-superconductor-junction device (e.g.~
Ref.~\citen{Alicea;Fisher:2011}), we may write a ``local''
topological number on the line as $M$, although the rigorous treatment 
requires a infinitely-extended spatial geometry (momentum preservation).
We calculate $M$ by the TKNN number in the system with chemical potential $\mu-V$, simply dropping the effects of the boundaries between the potential region and the others. 
In other words, the number $M$ is the TKNN number for the case when the potential width extends into the whole of the system. 
Although a one-dimensional class D superconductor is 
characterized by $\mathbb{Z}_{2}$\,\cite{Schnyder}, we use 
the TKNN number (i.e., $\mathbb{Z}$) in this paper 
since our aim is to compare the ``line-shaped'' area 
with the other area (e.g., the area on the left of the line 
potential) using the common index. 
Repeating a similar argument to the derivation of Eq.~(\ref{eq:tsc}),
we find that this number takes a non-zero value (i.e., $M=1$) when
\begin{align}
h > \sqrt{[4t - (\mu- V)]^{2} + \Delta^{2}}. \label{eq:1dtsc}
\end{align}
Let us divide the $h$-$V$ diagram into three regions,
using Eqs.~(\ref{eq:tsc}) and (\ref{eq:1dtsc}), as shown in
Fig.~\ref{fig:fig6}.  
The red solid line is defined by Eq.~(\ref{eq:tsc}), whereas the blue
dashed one is by Eq.~(\ref{eq:1dtsc}). 
We find that the latter line qualitatively
reproduces the line of the zero-energy bound states in
Fig.~\ref{fig:fig5}(a). 

Now, let us summarize phase diagram, as seen in Fig.~\ref{fig:fig6}(a). 
In phase I, the absence of the MGSs is found, since the system is
topologically trivial ($N = M = 0$). 
On the boundary between phases I and III, the energy gap in bulk
spectrum (\ref{eq:bulk}) closes. 
In phase II, there is no MGS. 
This comes from the fact that the topological number $N$ is the
same as $M$ (i.e., $N = M = 1$). 
On the boundary between phases II and III, 
the energy gap of the ``bulk'' spectrum in the effective 1D system
closes.  
In phase III, the appearance of the MGSs is found in the vicinity of the
potential wall, since the topological number $N$ is different from $M$
(i.e., $N=1$ and $M=0$). 
Taking the ultra-high potential limit in phase III leads to the
occurrence of the two zero-energy bound states, one of which is located
on the left side of the potential barrier, and the other of which is on
the right side. 
Reducing the potential height leads to the hybridization of these two
zero-energy states, with hopping proportional to $1/V$; as a result, the
degenerate zero eigenvalue is split. 
These arguments are sketched in Fig.~\ref{fig:fig6}(b). 

We remark that there is the difference between the 
numerical calculation [Fig.~\ref{fig:fig5}(a)] and the diagram in
Fig.~\ref{fig:fig8}(a). 
This difference comes from the fact that the latter diagram 
is obtained by the TKNN number in the two-dimensional 
system, not a genuine topological number 
in a one-dimensional (line) system. It is an interesting 
future issue to characterize our numerical results 
via a more definite topological number in a line system. 

\subsection{Non-zero-energy bound states in point-type potential}
\label{subsec:non-zero-point}
The absence of the zero-energy bound states is curious in the point-type
potential with ultra-high barrier ($V=100\,t$). 
Hu \textit{et al.}\cite{Hu;Liu:2013} discussed a link of non-zero bound
states around a non-magnetic impurity with the gapless surface modes,
using a finite-confinement effect in the chiral $p$-wave superfluid
surrounded by a disk-shaped rigid wall. 
Their arguments are interesting, but the map of the present model
to a chiral $p$-wave model is attainable when $h \to \infty$  
(We will show the derivation of the map in
Sec.\ref{subsec:low-energy}).  
Therefore, other ideas are desirable for explaining the difference from
zero energy and a link to the gapless surface modes in a finite magnetic
field. 

Let us consider this issue, from the viewpoint of continuous
deformation of the potential shape from a line to a point. 
We find the presence of the gapless surface modes in the
line-type potential with $V=100\,t$; the system is separated by the line
defect. 
Figure \ref{fig:fig4} shows that the bound states accommodate in the both
sides of the potential region. 
When one shortens the line length, the separated regions are connected
with each other. 
Then, interference would occur between the left and the right
zero-energy bound states. 
Thus, we speculate that the zero-energy bound states in the line defect 
would acquire non-zero energy in a point setting owing to the lack of
spatial separation. 
We will examine the validity of this idea in future. 


\subsection{Mixture of chiral $p$-wave and $s$-wave components}
\label{subsec:low-energy}
Let us derive an effective theory to understand the numerical results. 
Our approach is similar to a low-energy theory in the
semiconductor-superconductor junction systems,\cite{Alicea;Fisher:2011}
but we take a high-order correction to our description. 

We focus on the spatial uniform case. 
We seek the low-energy region for constructing the effective
theory, via the examination of the normal-part BdG Hamiltonian
[Eq.~(\ref{eq:bdg_hamiltonian}), with $\Delta =0$]. 
The energy eigenvalues in the particle subspace ($\tau^{3}=1$) are 
\(
E_{{\rm n},\veck}^{\pm} 
= 
\varepsilon_{\veck} 
\pm 
\sqrt{h^{2} + \alpha^{2}|\vecell_{\veck}|^{2}}
\).  
We focus on the case when the chemical potential is located at the
intermediate energy between the higher band and the lower band.
All the previous calculations in this article correspond to this case. 
Hence, the Fermi surface is determined by the solution of 
$0=E^{-}_{{\rm n},\veck}$. 
Since 
\(
E^{+}_{{\rm n},\veck}-E^{-}_{{\rm n},\veck} \ge 2h
\), 
the higher-energy contributions are negligible, when $h \to \infty$. 
A high magnetic-field value indicates that the electron spin is almost
polarized upward. 
Thus, the low-energy subspace is attainable by the spin-up projector 
\(
(1/2)(s^{0} + s^{3})\otimes \tau^{0}
\). 

The effective theory is built up by a perturbation approach, 
in the vicinity of the Fermi surface. 
The order of the perturbative expansions is evaluated by
$\alpha / h$ since the spin-orbit coupling mixes the low-energy sector
(spin up) with the high-energy sector (spin down). 
Also, the ratio of $|\Delta|$ to $h$ should be small since the
perturbation term contains the contributions of the pairing potential. 
Thus, the expansions are valid when $h \gg \alpha,\,|\Delta|$.  
Before the perturbative expansions, we perform a basis transformation to
take a higher-order correction. 
The procedure is similar to the Tani-Foldy-Wouthuysen
transformation~\cite{Itzykson;Zuber:2005} for the Dirac equation. 
The unitary-transformed BdG Hamiltonian is 
\(
\calH_{\veck,\eta} 
=
\calS_{\eta} \calH_{\veck} \calS^{-1}_{\eta}
\), 
with
\begin{equation}
 \calS_{\eta} 
= 
\exp \left[
i\frac{\eta}{2} s^{3} 
\otimes 
\left(
\frac{i\Delta}{|\Delta |}
\tau^{+} 
+
\text{h.c.} \right)
\right]. 
\label{eq:basis_transformation}
\end{equation}
We take so small angle $\eta$ that 
\(
\eta \sim \calO(\alpha /h)
\). 
In other words, we partially diagonalize the Hamiltonian, with the
restricted rotation $\calS$.  
We choose a free part of $\calH_{\veck,\eta}$, as seen in
Eq.~(\ref{eq:transformed_free}) in Appendix \ref{append:derivation}. 
Then, the 2nd-order Brillouin-Wigner approach~\cite{Hubac;Wilson:2010} leads
to an effective Hamiltonian, 
\begin{eqnarray}
\calH^{\rm eff}_{\veck}
&=&
 \frac{s^{0}+s^{3}}{2} \otimes 
\bigg[
\frac{\tau^{0} + \tau^{3}}{2} 
h^{\rm eff}_{\veck}
+
\frac{\tau^{0} - \tau^{3}}{2} 
(-h^{\rm eff}_{-\veck})^{\ast}
\nonumber \\
&&
+ 
( i \Delta^{\rm eff}_{\veck} \tau^{+} + \text{h.c.})
+ \calO(\alpha^{2}/h^{2})
\bigg], 
\label{eq:effective_hamiltonian}
\end{eqnarray} 
with 
\(
h^{\rm eff}_{\veck}
=
\varepsilon_{\veck} - h - (\alpha^{2} / h) |\vecell_{\veck}|^{2}
+
( |\Delta|^{2} / h) 
\) and 
\begin{equation}
\Delta^{\rm eff}_{\veck}
=
-\frac{2\alpha}{h} \Delta ( \ell_{2,\veck} + i\ell_{1,\veck})
-\frac{2 i \eta |\Delta|}{h} \Delta 
\label{eq:effective_gap}.
\end{equation} 
The detail calculations are shown in Appendix \ref{append:derivation},
but we concisely illustrate how effective gap function
(\ref{eq:effective_gap}) is produced from
the original $s$-wave mean-filed term. 
In Eq.~(\ref{eq:bdg_hamiltonian}), the spin-orbit couplings (the second
line) do not commute with the pairing potential term (the third line). 
Thus, in the perturbation analysis, the cross terms between them may
exist, and they lead to the effective superconducting gaps. 
When $\eta \to 0$ (i.e., no unitary transformation is performed), the
chiral $p$-wave low-energy behavior is survived, whereas the $s$-wave
contribution disappears. 
This result is consistent with the low-energy theory in the
semiconductor-superconductor junction systems.\cite{Alicea;Fisher:2011}  
Therefore, the full gap feature emerges as a higher-order correction in
the low-energy effective Hamiltonian. 
When a magnetic field is high, the chiral $p$-wave component is
predominant. 

A similar argument is possible for a 3D topological superconductor. 
The low-energy effective theory is attainable, taking
large-mass (nonrelativistic) limit.~\cite{Nagai;Machida:1404.1662} 
This assertion is reasonable, since the Zeeman magnetic field in the 2D
topological superconductor is regarded as the mass term. 
Moreover, we mention that, taking massless (ultrarelativistic)
limit\cite{Nagai;Machida:1404.1662,Michaeli} 
(i.e. a weak magnetic field in our system), the superconductivity is
robust against non-magnetic impurities, like $s$-wave superconductivity. 

The mixture of the chiral $p$-wave and $s$-wave characters in the
effective gap is interesting. 
It indicates that the robustness of the present model against
non-magnetic impurities can differ from typical behaviors in a chiral
$p$-wave model when the $s$-wave correction is negligible. 
We find that in the point setting (Sec.~\ref{sec:results2}) the system
is relatively robust around $h=1\,t$. 
Since we set $\alpha = 1\,t$ throughout this article, this robustness
can be related to the $s$-wave correction. 
It is an intriguing task to compare the present model with a chiral
$p$-wave model. 
The chiral $p$-wave superconductors also lead to the bound states with
almost zero energy around a non-magnetic impurity.\cite{Takigawa}
One can expect that the 2D topological $s$-wave superconductor
has a lot of behaviors similar to the ones in the chiral $p$-wave
superconductor. 
However, we guess that the quantitative differences are not
negligible in an intermediate magnetic field (e.g., $h \agt \alpha$)
since the map to chiral $p$-wave is attainable by perturbation with 
$h \gg \alpha$. 
Clarifying the difference from a chiral $p$-wave model
in the impurity effects is an important future issue. 

\section{Summary}
\label{sec:summary}
We studied how the MGSs appear in the 2D topological $s$-wave
superconductors with spatial inhomogeneity. 
Two kinds of the potential functions, line-type potential and point-type
one were examined. 
The line setting showed a definite link between the MGSs and the
gapless surface modes. 
The point setting showed that the quasiparticles with mid-gap energy
are much easily excited by impurities when the Zeeman
magnetic field increases. 
This result supports our previous
calculations\cite{Nagai;Machida:1312.3065} of the decrease of 
$T_{\rm c}$ by non-magnetic impurities. 
To understand the numerical results, we derived an effective theory
applicable to high magnetic fields. 
The effective gap was the mixture of the chiral $p$-wave and $s$-wave
components. 
The former is predominant when the magnetic field increases. 
Therefore, we claim that a chiral $p$-wave feature in the effective gap
function creates the MGSs. 

\section*{Acknowledgements}
We thank H. Nakamura, A. Shitade and K. Kobayashi for helpful discussions and comments. 
The calculations were performed using the supercomputing system PRIMERGY
BX900 at the Japan Atomic Energy Agency.  
This study has been partially supported by JSPS KAKENHI Grant Number 24340079, 26800197.

\appendix
\section{Derivation of an effective Hamiltonian}
\label{append:derivation}
We show the derivation of Eq.~(\ref{eq:effective_hamiltonian}). 
We start with the unitary-transformed BdG Hamiltonian 
\(
\calH_{\veck,\eta} 
=
\calS_{\eta} \calH_{\veck} \calS^{-1}_{\eta}
\), where $\calS_{\eta}$ is defined by
Eq.~(\ref{eq:basis_transformation}). 
We take so small angle $\eta$ that 
\(
\eta \sim \calO(\alpha /h)
\). 
After straightforward calculations, we find that 
\(
\calH_{\veck,\eta} 
=
\calH^{(0)}_{\veck,\eta}
+
\calV_{\veck,\eta}
+
\calO(\eta^{2})
\), 
with
\begin{eqnarray}
\calH^{(0)}_{\veck,\eta}
&=&
\frac{s^{0}+s^{3}}{2} \otimes [
(\varepsilon_{\veck} - h)\tau^{3} 
+(i \Delta^{(+)}_{\veck, \eta}\tau^{+}+ \text{h.c.})]
\nonumber \\
&&
+
\frac{s^{0}-s^{3}}{2} \otimes [
(\varepsilon_{\veck} + h)\tau^{3} 
+(i \Delta^{(-)}_{\veck, \eta}\tau^{+}+ \text{h.c.})
], 
\nonumber \\
&&
\label{eq:transformed_free}
\\
\calV_{\veck,\eta}
&=&
s^{1} \otimes 
( \alpha \ell_{1,\veck} + i \Delta^{(1)}_{\eta}) \tau^{0}
\nonumber \\
&&
+
s^{2} \otimes [ \alpha \ell_{2,\veck} \tau^{3}
+
(
i \Delta^{(2)}_{\veck,\eta} \tau^{+} + \text{h.c.}
)].
\label{eq:transformed_perturb}
\end{eqnarray}
The four additional gap functions are 
\mbox{
\(
\Delta^{(\pm)}_{\veck, \eta}
=
\mp i \eta (\varepsilon_{\veck} \mp h) ( \Delta / |\Delta|)
\)},
\mbox{
\(
\Delta^{(1)}_{\eta}
=
-i \eta |\Delta |
\)
}, and 
\mbox{
\(
\Delta^{(2)}_{\veck, \eta}
=
[ 
1 -(\eta \alpha \ell_{1,\veck}/|\Delta|)
]\Delta
\)}. 
We set the free part for the perturbation approach as
$\calH^{(0)}_{\veck,\eta}$. 

Now, we derive the low-energy effective Hamiltonian.
The 2nd-order Brillouin-Wigner approach~\cite{Hubac;Wilson:2010} leads
to an effective Hamiltonian 
\begin{equation}
 \calH^{\rm eff}_{\veck}
= 
\calP \calH^{(0)}_{\veck,\eta} \calP 
+
\sum_{m=1}^{2}
(\calP \calV_{\veck,\eta} \calQ )
\calR_{\veck,\eta,m}
(\calQ \calV_{\veck,\eta} \calP ),
\label{eq:2nd_BWA}
\end{equation}
with
\(
\calP = (1/2)(s^{0}+s^{3})\otimes \tau^{0}
\), 
\(
\calQ = s^{0}\otimes \tau^{0} - \calP
\)
and  
\(
\calR_{\veck,\eta,m}
=
\calQ
[
E^{(0)}_{\veck, \eta, m} - \calQ \calH^{(0)}_{\veck,\eta}\calQ 
]^{-1}
\). 
The non-perturbative energy 
\(
E^{(0)}_{\veck, m}
\) 
is related to the eigenvalues of $\calP \calH^{(0)}_{\veck,\eta}\calP$: 
\(
\calP \calH^{(0)}_{\veck,\eta} \calP
=
(1/2)(s^{0}+s^{3})\otimes 
{\rm diag}\{E^{(0)}_{\veck,\eta,1},\,E^{(0)}_{\veck,\eta,2}\}
\). 
We approximate $\calR_{\veck,\eta,m}$ by the value on the Fermi surface, 
\(
\calR_{\veck,\eta,m} \approx \calR_{\veck_{\rm F},\eta,m}
\), 
with the Fermi momentum $\veck_{\rm F}$ given by 
\(
\epsilon_{\veck_{\rm F}} = h + \calO(\alpha^{2} / h^{2})
\). 
Taking the leading term,\cite{note1} we find that 
\(
R_{\veck,\eta,m}
\approx 
(1/2)(s^{0}-s^{3})\otimes (-1/2h)\tau^{3}
\). 
Therefore, performing straightforward algebraic calculations about the
Pauli matrices, we obtain Eq.~(\ref{eq:effective_hamiltonian}). 
In the effective gap $\Delta^{\rm eff}_{\veck}$, the chiral $p$-wave
component (the first term) comes from $\Delta^{(2)}_{\veck,\eta}$. 
In contrast, the $s$-wave component (the second term) is attributable to
$\Delta^{(1)}_{\eta,\veck}$. 
We remark that the contribution from $\Delta^{(+)}_{\veck,\eta}$
in Eq.~(\ref{eq:transformed_free}) is negligible, since on the Fermi
surface  
\(
\Delta^{(+)}_{\veck,\eta}
\sim 
\eta \calO(\alpha^{2}/h^{2})
\sim 
\calO(\alpha^{3}/h^{3})
\). 
The contribution from $\Delta^{(-)}_{\veck,\eta}$ is irrelevant to the
low-energy sector, owing to the spin-up projector $\calP$.


\begin{thebibliography}{99}
\bibitem{Anderson}
P. W. Anderson, 
J. Phys. Chem. Solids, \textbf{11}, 26 (1959). 
\bibitem{Tsuneto:1962}
T. Tsuneto, 
Prog. Theore. Phys. \textbf{28}, 857 (1962). 
\bibitem{deGennes:1999}
P. G. de Gennes, 
\textit{Superconductivity of Metals and Alloys} 
(Westview Press, Boulder, 1999) Chap.8. 
\bibitem{Kopnin:2001}
N. B. Kopnin, 
\textit{Theory of Nonequilibrium Superconductivity} 
(Oxford University Press, New York, 2001). 
\bibitem{Balatsky}
A. V. Balatsky, I. Vekhter, and J.-X. Zhu, 
Rev. Mod. Phys. {\bf 78}, 373 (2006). 
\bibitem{Hirschfeld;Woelfe:1986}
P. Hirschfeld, D. Vollhardt, and P. W\"{o}lfle, 
Solid State Commun \textbf{59}, 111 (1986). 
\bibitem{SchmittRink;Varma:1986}
S. Schmitt--Rink, K. Miyake, and C. M. Varma, 
Phys. Rev. Lett. \textbf{57}, 2575 (1986).
\bibitem{Hotta:1993}
T. Hotta, 
J. Phys. Soc. Jpn. \textbf{62}, 274 (1993). 
\bibitem{Preosti;Muzikar:1996}
G. Preosti and P. Muzikar, 
Phys. Rev. B \textbf{54}, 3489 (1996). 
\bibitem{Maki;Puchkaryov:1999}
K. Maki and E. Puchkaryov, 
Europhys. Lett. \textbf{45}, 263 (1999). 
\bibitem{Maki;Puchkaryov:2000}
K. Maki and E. Puchkaryov, 
Europhys. Lett. \textbf{50}, 533 (2000). 
\bibitem{Fu;Berg:2009}
L. Fu and E. Berg, 
Phys. Rev. Lett. \textbf{105}, 097001 (2010).
\bibitem{Sato:2010}
M. Sato, 
Phys. Rev. B \textbf{81}, 220504(R) (2010). 
\bibitem{Sato;Fujimoto:2009}
M. Sato, Y. Takahashi, and S. Fujimoto, 
Phys. Rev. Lett. \textbf{103}, 020401 (2009). 
\bibitem{Sato;Fujimoto:2010}
M. Sato, Y. Takahashi, and S. Fujimoto, 
Phys. Rev. B \textbf{82}, 134521 (2010).
\bibitem{Fradkin:2013}
E. Fradkin, 
\textit{Field Theories of Condensed Matter Physics} 
(Cambridge University Press, Cambridge, UK, 2013) Chap.15. 
\bibitem{Read;Green:2000}
N. Read and D. Green, 
Phys. Rev. B \textbf{61}, 10267 (2000).
\bibitem{Teo;Kane:2010}
J. C. Y. Teo and C. L. Kane,  
Phys. Rev. B \textbf{82}, 115120 (2010).
\bibitem{Nishida:2010}
Y. Nishida, 
Phys. Rev. D \textbf{81} 074004 (2010).
\bibitem{Alicea;Fisher:2011}
J. Alicea, Y. Oreg, G. Refael, F. von Oppen, and M. P. A. Fisher, 
Nat. Phys. \textbf{7}, 412 (2011).
\bibitem{Ohashi}
Y. Ohashi, 
Phys. Rev. A \textbf{83}, 063611 (2011). 
\bibitem{Sau;Demler:2013}
J. D. Sau and E. Demler, 
Phys. Rev. B \textbf{88}, 205402 (2013). 
\bibitem{Ichioka;Machida:2007}
M. Ichioka and K. Machida, 
Phys. Rev. B \textbf{76}, 064502 (2007). 
\bibitem{Samokhin}
K. V. Samokhin, 
in 
\textit{
Non-Centrosymmetric Superconductors: Introduction and Overview}, 
ed. E. Bauer and M. Sigrist 
(Springer, Heidelberg, 2012) p.269.
\bibitem{Hu;Liu:2013}
H. Hu, L. Jiang, H. Pu, Y. Chen, and X.-J. Liu, 
Phys. Rev. Lett. \textbf{110}, 020401 (2013). 
\bibitem{Thouless;Nijs:1982}
D. J. Thouless, M. Kohmoto, M. P. Nightingale, and M. den Nijs, 
Phys. Rev. Lett. \textbf{ 49}, 405 (1982). 
\bibitem{Nagai;Machida:1312.3065}
Y. Nagai, Y. Ota, and M. Machida, J. Phys. Soc. Jpn., {\bf 83}, 094722 (2014). 
\bibitem{Sakurai;Sugiura:2003}
T. Sakurai and H. Sugiura, 
J. Comput. Appl. Math. \textbf{159}, 119 (2003). 
\bibitem{Nagai;Sakurai:2013}
Y. Nagai, Y. Shinohara, Y. Futamura, Y. Ota, and T. Sakurai, 
J. Phys. Soc. Jpn. \textbf{82}, 094701 (2013).
\bibitem{Itzykson;Zuber:2005}
C. Itzykson and J.-B. Zuber, 
\textit{Quantum Field Theory} 
(Dover, New York, 2005).
\bibitem{Hubac;Wilson:2010}
I. Huba\v{c} and S. Wilson, 
\textit{Brillouin-Wigner Methods for Many-Body Systems} 
(Springer, Heidelberg, 2010). 
\bibitem{note1}
A definite formula is 
\(
\calR_{\veck_{\rm F},\eta,m}
=
(1/2)(s^{0}-s^{3})\otimes 
[
(-1/2h)\tau^{3}
+
(-\eta / 2h|\Delta|)
(i\Delta \tau^{+} + \text{h.c.})
+
\calO(\eta^{2})]
\). 
The correction from the second term in the above formula to
	Eq.~(\ref{eq:effective_hamiltonian}) is 
	an overall (non-zero) complex-number constant in
	the $s$-wave contribution. 
\bibitem{Schnyder}
A. P. Schnyder, S. Ryu, A. Furusaki, and A. W. W. Ludwig, 
Phys. Rev. B \textbf{78}, 195125 (2008). 
\bibitem{Nagai;Machida:1404.1662}
Y. Nagai, Y. Ota, and M. Machida, 
Phys. Rev. B \textbf{89}, 214506 (2014).
\bibitem{Michaeli}
K. Michaeli and L. Fu, Phys. Rev. Lett. \textbf{109}, 187003 (2012). 
\bibitem{Takigawa}
M. Takigawa, M. Ichioka, K. Kuroki, and Y. Tanaka, 
Phys. Rev. B \textbf{72}, 224501 (2005). 
\end{thebibliography}
\end{document}